\documentclass[%
reprint,
superscriptaddress,
amsmath,
amssymb,
pra,
]{revtex4-1}

\usepackage{comment}
\usepackage{graphicx}
\usepackage{xcolor}
\usepackage{dcolumn}
\usepackage{bm}
\usepackage{braket}
\usepackage[colorlinks=true, linkcolor=red, anchorcolor=black, citecolor=blue,
urlcolor = blue]{hyperref}

\begin{document}

\preprint{APS/123-QED}

\title{Zeeman-resolved Autler-Townes splitting in Rydberg atoms with tunable resonances and a single transition dipole moment}

\author{Noah Schlossberger}
 \affiliation{Department of Physics, University of Colorado, Boulder, Colorado 80302, USA}
 \email{noah.schlossberger@nist.gov}
\affiliation{Associate of the National Institute of Standards and Technology, Boulder, Colorado 80305, USA}
\author{Andrew P. Rotunno}
\affiliation{ National Institute of Standards and Technology, Boulder, Colorado 80305, USA}
\author{Aly Artusio-Glimpse}
\affiliation{ National Institute of Standards and Technology, Boulder, Colorado 80305, USA}
\author{Nikunjkumar Prajapati}
\affiliation{ National Institute of Standards and Technology, Boulder, Colorado 80305, USA}
\author{Samuel Berweger}
\affiliation{ National Institute of Standards and Technology, Boulder, Colorado 80305, USA}
\author{Dangka Shylla}
 \affiliation{Department of Physics, University of Colorado, Boulder, Colorado 80302, USA}
\affiliation{Associate of the National Institute of Standards and Technology, Boulder, Colorado 80305, USA}
\author{Matthew T. Simons}
\affiliation{ National Institute of Standards and Technology, Boulder, Colorado 80305, USA}
\author{Christopher L. Holloway}
\affiliation{ National Institute of Standards and Technology, Boulder, Colorado 80305, USA}
 
\date{\today}

\begin{abstract}
Applying a magnetic field as a method for tuning the frequency of Autler-Townes splitting for Rydberg electrometry has recently been demonstrated. 
In this paper we provide a theoretical understanding of EIT signals in the presence of a large magnetic field, as well as demonstrate some advantages of this technique over traditional Autler-Townes based electrometry.
We show that a strong magnetic field provides a well-defined quantization axis regardless of the optical field polarizations, 
we demonstrate that by separating the $m_J$ levels of the Rydberg state we can perform an Autler-Townes splitting with a single participating dipole moment, 
and we demonstrate recovery of signal strength by populating a single $m_J$ level using circularly polarized light.
\end{abstract}

\maketitle

\section{\label{sec:intro} Introduction}

Highly-excited atomic Rydberg states have been demonstrated as `self-calibrated' microwave field sensors \cite{sedlacek2012microwave, holloway2014aps, 9748947, holloway2017electric, meyer2020assessment, fancher2021rydberg, yuan2023quantum, liu2023electric} due to strong calculable \cite{vsibalic2017arc} dipole matrix elements between Rydberg states. 
States are experimentally populated and probed by multi-photon electromagnetically induced transparency (EIT) \cite{mohapatra2007coherent}, and field measurements are made by interpreting the laser absorption spectrum through an atomic vapor cell. 
Since each atomic species has a defined set of discrete transition frequencies, the ability to tune these transitions for measuring arbitrary microwave frequency 
has been demonstrated using 
off-resonant measurements~\cite{simons2016using, PhysRevA.104.032824, 10.1063/1.4996234, hu2022continuously, meyer2021waveguide},
DC electric~\cite{osterwalder1999using, ma2022measurement, holloway2022electromagnetically, duspayev2023high},
AC electric~\cite{holloway2022electromagnetically, noel1998frequency, bohlouli2007enhancement, berweger2023rydberg, rotunno2023detection}, 
and as we use here, DC magnetic fields~\cite{bao2016polarization,naber2017electromagnetically,shi2023electric,li2023magnetic}.
Experimental Autler-Townes (AT) measurements are often contaminated by state mixing due to laser and microwave field polarization~\cite{sedlacek2013atom,wang2023precise} and the strength~\cite{r13, r12}
and spatial inhomogeneity~\cite{rotunno2023investigating} of the microwave fields. 

There are several key benefits to working in a strong magnetic field, i.e. a field such that $\mu_B B \gg \Omega$ where $\mu_B$ is the Bohr magneton, $B$ is the applied magnetic field strength, and $\Omega$ are the larger of the Rabi rates of the probe and coupling lasers. In this regime (of order mT), the magnetic field defines the quantization axis. In comparison to a Hamiltonian consisting of only photon interactions, this situation is much less ambiguous (generally when $\mu_B B$ is small compared to the Rabi rates, a Rabi rate much stronger than the other Rabi rates can unambiguously set the quantization axis, e.g. the radio frequency (RF) photon in AT).

Another key benefit of working in the large $B$ regime is that individual $m_J$ levels of the Rydberg state in the EIT signal can be resolved. This allows angular momentum dynamics to be directly observed, a valuable tool for polarization studies. Another useful feature of an isolated $m_J$ level is that for a well-defined RF polarization, there is only a single possible transition for an Autler Townes splitting, meaning that electrometry can be calibrated using a single unambiguous transition dipole matrix element. 
In this paper, we will add theoretical foundation to the results presented in, e.g., \cite{shi2023electric, li2023magnetic}, 
as well as illustrate interesting aspects of using stretch states ($m_J=J$) on the atomic spectra, including increase in signal-to-noise when using circularly polarized light,
and ramifications for calibrating electric-field sensors based on EIT-AT.

\section{EIT in the presence of a strong magnetic field}
\begin{figure}[h!]
    \includegraphics[scale = .8]{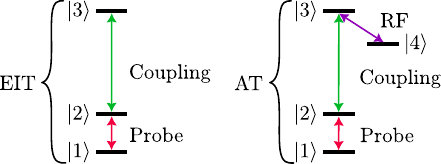}
    \caption{Energy level schemes for EIT and Rydberg AT.}
    \label{fig:labels}
\end{figure}

The energy level diagram for a two-photon cascade EIT scheme and the RF extension for Rydberg Autler-Townes splitting is shown in Figure~\ref{fig:labels}.

State $\ket{1}$ is the ground $S_{1/2}$ state's highest hyperfine manifold, state $\ket{2}$ is the $P_{3/2}$ state's highest hyperfine manifold, state $\ket{3}$ is a Rydberg $nD_{5/2}$ state, and $\ket{4}$ is a nearby Rydberg state.

A theoretical approach to treating the relevant states in an EIT or Autler-Townes scheme involves treating $m_F$ as the quantum number describing the angular momentum in the low-principal quantum number states where $\mu_B B$ is small compared to the hyperfine energy splitting ($\sim$GHz in $|1\rangle$ and $\sim$100s of MHz in $|2\rangle$), and using $m_J$ as the quantum number describing the angular momentum of the Rydberg states, where $\mu_B B$ is large compared to the hyperfine energy splitting and the hyperfine structure cannot be resolved. This treatment is sufficient because the $m_I$ dependence of the Zeeman shift is negligible~\cite{ZeemanDipoleTreatment}. Note that for a large enough field, the intermediate $|2\rangle$ state will have similar strengths of hyperfine and Zeeman splittings, meaning neither treatment is sufficient~\cite{naber2017electromagnetically}.
The two pictures can be linked with selection rules acting on $m_F$ in the lower states and $m_I + m_J$ in the Rydberg states, and summing over all $m_I$.

In the presence of a magnetic field, each level undergoes a shift in energy of
\begin{equation}
\Delta f = \frac{\mu_B B}{h} \cdot
\begin{cases}
    g_F m_F& \textrm{low-}n\textrm{ states}\\
    g_J m_J& \textrm{Rydberg states}
\end{cases},
\end{equation}
where $g_J$ and $g_F$ are the Land\'e g-factors as described in \cite{steck2003cesium}.

The transition dipole matrix element $d_{m_{F2} \rightarrow m_{J3}}$ from a hyperfine $m_F$ state of $\ket{2}$ to a fine structure $m_J$ state of $\ket{3}$ is calculated by projecting the fine structure state into its associated hyperfine states and summing over the dipole matrix element from each component:
\begin{multline}
   d_{m_{F2} \rightarrow m_{J3}} \equiv \langle n_2, L_2, J_2, F_2, m_{F2} | e \vec{r} | n_3, L_3, J_3, m_{J3}\rangle \\
    =
    \sum_{F'}\sum_{m_F' = -F'}^{F'}\sum_{m_I' = -I}^{I} C^{F' m_F'}_{I m_I' J_3 m_{J3}} \cdot
    \\
    \langle n_2, L_2, J_2, F_2, m_{F2} | e \vec{r} | n_3, L_3, J_3, F', m_{F}'\rangle ,
\end{multline}
where $C^{J M}_{j_1 m_1 j_2 m_2}$ are the Clebsch-Gordan coefficients, $n$ is the principle quantum number, $L$ is the orbital angular momentum quantum number, $J$ is the total electronic angular momentum quantum number, $F$ is total atom angular momentum quantum number, $m_{F}$ and $m_J$ are the projection of the the total electronic and total angular momenta respectively onto the quantization axis, and the number in the subscripts correspond to the state as labeled in Figure~\ref{fig:labels}.
Neglecting saturation effects, the relative total transition strength $P(m_J)$ going to any different $m_J$ level in the Rydberg state is proportional to
\begin{equation}
    P(m_J) \propto \sum_{m_{F1}}\sum_{m_{F2}} d_{m_{F1} \rightarrow m_{F2}}^2 d_{m_{F2}\rightarrow m_J}^2. \label{eq:transitionstrengthestimator}
\end{equation}
The relative intensities of EIT peaks are approximately described by these probabilities, with complication arising from spontaneous emission and cycling rates of the transitions. The theoretical transition dipole moments can be calculated using the ARC python package \cite{vsibalic2017arc}. The dipoles are calculated without Zeeman or Stark effects, which generally adjust atomic wave functions and therefore the associated transition dipole matrix elements \cite{sedlacek2012microwave}.

\begin{figure}[h!]
\includegraphics[width = \linewidth]{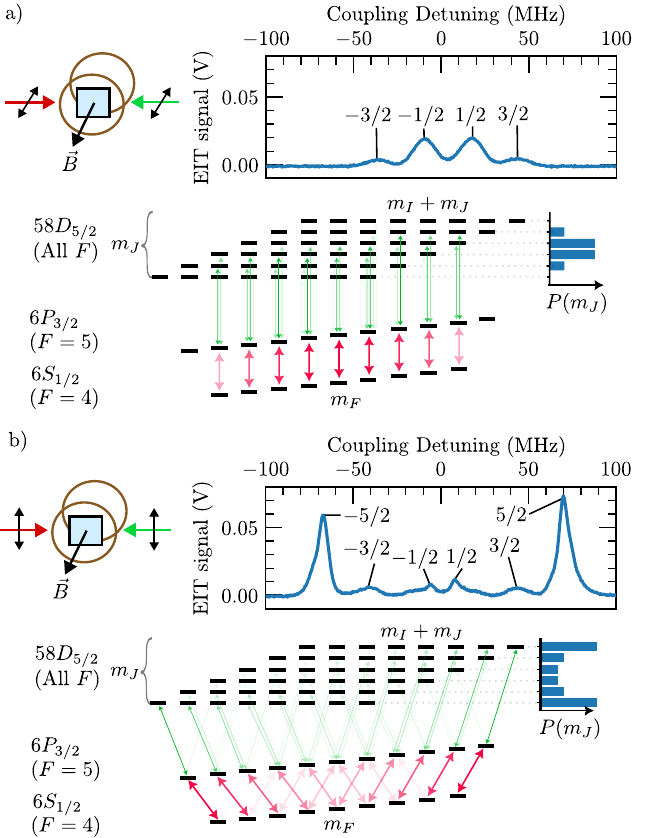}
\caption{Measured EIT signals in the presence of a strong (1.85(1)~mT) magnetic field either a) aligned with or b) orthogonal to the polarization axis. For each case, the polarization and magnetic field axes are visualized in the top left, and the resulting EIT signal is shown in the top right. The energy level diagrams show the angular momentum pathways from the ground state to the various $m_J$ levels, with the opacity of each transition scaled by the square of its transition dipole moment. To the right of the diagram are theoretical signal strengths $P(m_J)$ given by Equation \ref{eq:transitionstrengthestimator}.} \label{fig:EnergyLevelDiagrams}
\end{figure}

An EIT signal is taken on the ${6S_{1/2} (F=4)} \rightarrow {6P_{3/2} (F=5)} \rightarrow {58D_{5/2}}$ pathway in Cs in the presence of a magnetic field in Figure~\ref{fig:EnergyLevelDiagrams}.
We use a simple experimental setup~\cite{holloway2014aps} with counter-propagating probe and coupling beams. 
We detect laser absorption using differential detection, subtracting a reference probe-only beam transmittance from the coupling-overlapped EIT transmittance on a balanced photodetector. The coupling laser detuning is scanned many MHz around the atomic state resonance while the probe is locked on the $\ket{1} \rightarrow \ket{2}$ transition. 
A simultaneous field-free vapor cell measurement provides shot-to-shot calibration of offset and scaling for the detuning axis.

The relative orientation of the magnetic field with respect to the laser orientations yields a very different distribution of $m_J$ peak strengths in the EIT signal. This demonstrates that the magnetic field sets the quantization axis, such that when the polarizations of the lasers are aligned with the magnetic field they are $\pi$ transitions and when they are perpendicular they become a superposition of $\sigma^+$ and $\sigma^-$ polarizations when projected onto the quantization axis. This, combined with the fact that for a given initial state the dipole moments for angular momentum changing transitions are stronger when they bring the angular momentum towards the nearest extrema means that the stretch states $m_J = \pm 5/2$ give the strongest EIT signal in this case.

In the perpendicular case, the $m_J$ peaks appear to have internal structure leading to  unequal frequency spacing, and the $m_J = \pm 5/2, \pm 1/2$ peaks are narrower than the $m_J = \pm 3/2$ peaks. This can be understood by considering EIT pathways and population dynamics. Each possible transition pathway that shows up in the EIT signal has its own coupling laser frequency at which EIT will occur, which depends on the Zeeman shifts on all three associated states:
\begin{multline}
    \Delta f (m_{F1}, m_{F2}, m_{J3}) =\\ \frac{\mu_B B}{h}\big(\underbrace{g_{J_3} m_{J3} - g_{F_2} m_{F2}}_\textrm{Coupling Shift} + \underbrace{\frac{f_\textrm{c}}{f_\textrm{p}}(g_{F_2} m_{F2} - g_{F_1}m_{F1})}_\textrm{Probe Doppler Shift}\big) \label{eq:deltaF},
\end{multline}
where $f_\textrm{p}$ and $f_\textrm{c}$ are the frequencies of the probe and coupling lasers respectively.

We can then theoretically construct the EIT signal by summing a finite-width Gaussian for each transition with its amplitude prescribed by the transition strength given by the product of the square of the transition dipole moments and its location prescribed by the $\Delta f$ in Equation \ref{eq:deltaF}. This can be written as
\begin{multline}
    V(f) = \sum_{m_{F1}} \sum_{m_{F2}} \sum_{m_{J3}}  d_{m_{F1} \rightarrow m_{F2}}^2 d_{m_{F2}\rightarrow m_J}^2 \\
    e^{\frac{-\left(f-\Delta f (m_{F1}, m_{F2}, m_{J3}) \right)^2}{2 \sigma^2}}, \label{eq:vfsimp}
\end{multline}
where $\sigma$ is the empirical Gaussian width of each individual transition pathway.

\begin{figure}[h!]
    \includegraphics[width = .95\linewidth]{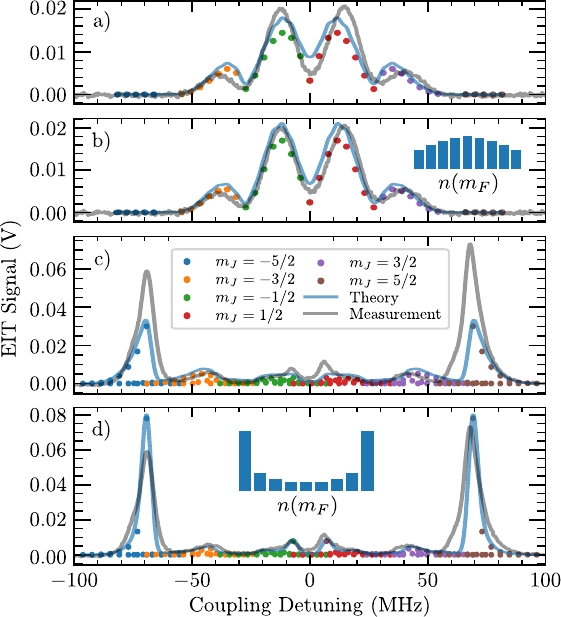}
    \caption{Theoretical EIT signals for Cs in the presence of a 1.85(1)~mT magnetic field for light polarizations aligned (a,b) and perpendicular to (c,d) to the magnetic field. a) and c) present the simplified theory from Equation \ref{eq:vfsimp} which only takes into account dipole moments and frequency shifts of each pathway while b) and d) include population weighting as in Equation \ref{eq:vf}. The only floating parameters are an overall amplitude scaling and the width $\sigma$ of each transition, which is common among all four theory curves.} 
    \label{fig:TheoryEIT}
\end{figure}

Further realism can be added by putting the laser powers and decay rates into a model to yield the steady state relative $m_F$ populations $\{n(m_{F1})\}$ in the $6S_{1/2} (F=4)$ or $\ket{1}$ ground state, and weighting each transition by the relative population:
\begin{multline}
    V(f) = \sum_{m_{F1}} \sum_{m_{F2}} \sum_{m_{J3}}  n(m_{F1}) d_{m_{F1} \rightarrow m_{F2}}^2 d_{m_{F2}\rightarrow m_J}^2 \\
    e^{\frac{-\left(f-\Delta f (m_{F1}, m_{F2}, m_{J3}) \right)^2}{2 \sigma^2}} \label{eq:vf}.
\end{multline}
Calculations for both polarization cases are shown in Figure~\ref{fig:TheoryEIT}. 
Populations of the ground state are calculated using RydIQule~\cite{miller2023rydiqule} to numerically solve a many-state model. 
The model used the Zeeman basis of \{$m_F, m_F, m_J$\} for the \{$6S_{1/2}, 6P_{3/2},58D_{5/2}$\} states respectively, summing over $m_I$, and accounted for Rabi scaling and decay branching using Zeeman resolved dipole moments~\cite{vsibalic2017arc}, as well as a transit decay from the Rydberg state to all $m_F$ sub-levels of $\ket{1}$ equally.
The laser Rabi frequency scaling used in calculating Figure~\ref{fig:TheoryEIT}~b) are a probe field of 0.1~MRad/s/$ea_0$ and a coupling field of 20~MRad/s/$ea_0$, and the laser electric fields used in calculating Figure~\ref{fig:TheoryEIT}~d) are a probe field of 0.1~MRad/s/$ea_0$, and coupling field of 5~MRad/s/$ea_0$.

The dipole moments alone are able to capture the general shape of the waveforms, while adding in the $n(m_F)$ weights captures the fact that with spontaneous emission, $\pi$ polarized light generally moves the population towards lower absolute angular momentum while $\sigma^+ + \sigma^-$ polarized light generally moves the population towards higher absolute angular momentum states.

Generally, this theory explains the shapes of each $m_J$ peak in the EIT signal in the presence of a strong magnetic field and indicates that each peak corresponds predominantly to a pure $m_J$ level in the Rydberg state. It is also notable that in the parallel polarization case, the apparent width of each $m_J$ peak comes predominantly from Zeeman-induced spread in the resonant coupling frequencies (Equation \ref{eq:deltaF}) for each pathway (see Figure~\ref{fig:TheoryEIT}~b).
\section{Autler-Townes}
To find the shift in the microwave frequency at which resonant Autler-Townes splitting will occur, one can simply subtract the Zeeman shifts on the $m_J$ levels of the two Rydberg states:
\begin{equation}
    \Delta f = (g_{J4} m_{J4} - g_{J3} m_{J3}) \frac{\mu_B B}{h} \label{eq:delftaf}.
\end{equation}
It is worth noting that if the fourth state is lower in energy than the third state, this apparent $\Delta f$ changes sign, since the microwave resonance corresponds to the \emph{magnitude} of the energy difference. For the stretch transition from a $(n) D_{5/2}$ to a lower energy $(n+1) P_{3/2}$ state, the shift in the frequency at which the Autler-Townes is sensitive to is given by 
\begin{equation}
    \Delta f = -\left(  g_{P_{3/2}}\frac{3}{2} -  g_{D_{5/2}}\frac{5}{2}\right)\frac{\mu_B B}{h}
    = \frac{\mu_B B}{h},
\end{equation}
which gives a shift of 13.996~MHz per~mT of magnetic field. 
Note that the difference of $m_J g_J$ products simplifies to 1 for a stretch state transition. 

\begin{figure}[h!]
    \includegraphics[width = .9\linewidth]{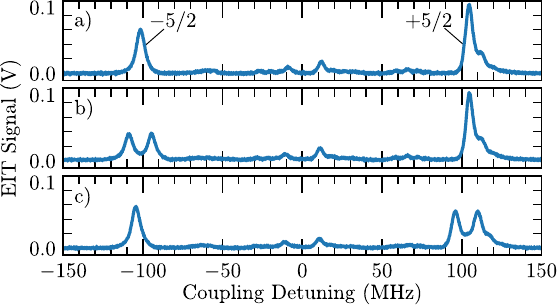}
    \caption{Measured Autler-Townes splittings in individual $m_J$ levels via the ${58D_{5/2}(m_J = \pm 5/2)} \rightarrow {59 P_{3/2} (m_J = \pm 3/2)}$ transitions of Cs in the presence of 2.78(1)~mT. a): No RF is applied. b): An RF frequency of 3.5426~GHz is applied. c): An RF frequency of 3.6205~GHz is applied. The frequency difference of 77.9~MHz between b) and c) corresponds to 2$\mu_B B$.} \label{fig:indsplitting}
\end{figure}

Using the optical fields perpendicular to the magnetic field populates the $m_J = \pm 5/2$ states of the Rydberg level, in which the Zeeman-induced shift on the Autler-Townes frequency is maximized. In this configuration, we can independently split the $m_J = -5/2$ and $+5/2$ levels at different microwave frequencies, as demonstrated in Figure~\ref{fig:indsplitting}.

A key benefit of lifting the degeneracy of the $m_J$ levels of the Rydberg state is that Autler-Townes splittings occur with only a single dipole moment. 
The typical configuration \cite{holloway2014aps} for Autler-Townes electrometry uses linearly polarized fields with the polarization axes of both optical and microwave photons all aligned, such that each transition is a $\pi$ transition in the strong microwave field's quantization axis. 
However, even with pure polarization this can populate multiple $m_J$ levels (the angular momentum from the nucleus can couple into the electron during the projection from $m_F$ onto $m_J$), so there will be multiple Autler-Townes splittings each with a different dipole moment. 
Calibration of the splitting to electric field amplitude is obscured by the fact that several splittings with different transition dipole moments are occurring simultaneously, with relative amplitudes depending on steady-state population dynamics and therefore nontrivial to estimate. 
For a $D_{5/2} \rightarrow P_{3/2}$ transition, there are two unique dipole elements to account for for $\pi$ microwaves ($m_J = {\pm 1/2 \leftrightarrow \pm 1/2}$, ${\pm 3/2 \leftrightarrow \pm 3/2}$  ) and four more for unintended $\sigma$ transitions, 
while for a $D_{5/2} \rightarrow F_{7/2}$ transition there are four unique dipoles for $\pi$ transitions and six more for unintended $\sigma$ transitions.

\begin{figure}[h!]
\includegraphics[width = \linewidth]{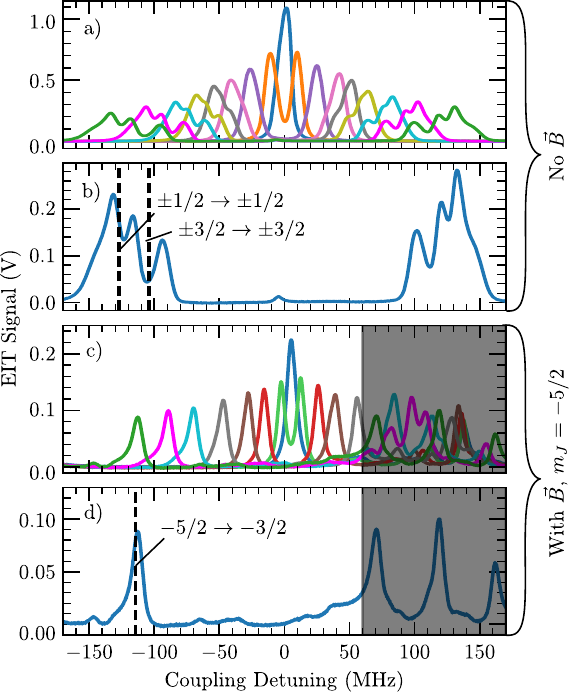}
\caption{Measured Autler-Townes splittings on the Cs $58D_{5/2} \rightarrow 59P_{3/2}$ transition with and without $m_J$ selectivity for various RF fields up to 3.08 V/m. a,b): No magnetic field is applied, and optical and RF fields share the same polarization axis.  c,d): 3.24(1)~mT of magnetic field is applied orthogonal to the RF and optical polarization axis, and the $m_J = -5/2$ hyperfine state is populated and split, taking the corresponding peak location to be 0 detuning. The area in which other $m_J$ levels start to interfere with the right Autler-Townes peak at high RF fields is greyed out. b) and d) show the EIT signals with 3.08 V/m of RF applied. The field strength is found by fitting the peak location of the $-5/2 \rightarrow -3/2$ transition vs the square root of the power applied to the horn in the low field regime (scaling the effective field by $1/\sqrt{2}$ to account for projecting the RF into the quantization axis) and the theoretical peak locations for b) and d) are then calculated using the relative dipole moments of each transition and represented with a dashed vertical line. Note that precise dipole moment comparisons are obscured by frequency dependent field amplitudes and atomic high-field effects in this regime.}
\label{fig:SingleDipole}
\end{figure}

In contrast, applying a magnetic field large enough to separate the $m_J$ levels means that when Autler-Townes is performed, the number of transitions reduces to one for pure microwave polarization and at most two more for mixed polarization, and these will be off-resonant when tuned to a particular $m_J \rightarrow m_{J'}$ transition.
A particularly clean transition is the $D_{5/2} (m_J=5/2)\rightarrow P_{3/2} (m_J=3/2)$ stretch state transition, in which only a single transition is allowed from the initial state. This is demonstrated in Figure~\ref{fig:SingleDipole}.

In the no-magnetic-field case with aligned linearly polarized light (the standard case for Rydberg AT electrometry), the signal is difficult to interpret, but it is clear that the AT peaks spread out and have structure as the microwave field increases amplitude, while the $m_J$-resolved case stays narrow at high-field, indicating it is being split by a single transition.

This makes isolating $m_J$ levels with a magnetic field an appealing option for precision electrometry using Rydberg atom sensors, with the potential drawbacks being interference of the microwave field by the geometry of the magnetic field sources, as well as potential reduction in signal strength, which can be avoided with circularly polarized light, as demonstrated in the next section.

\section{Polarization}
In Figure~\ref{fig:EnergyLevelDiagrams}, we saw that different linear polarizations lead to EIT in different $m_J$ levels, with the most useful case being polarization perpendicular to the magnetic field direction because it excites the stretch states, which are most separated in terms of magnetic field sensitivity. However, each $m_J$ peak is significantly weaker than the standard EIT peak without a magnetic field, since the population is split between the $+5/2$ and $-5/2$ states. This signal reduction ultimately reduces the minimum detectable field using this technique.

\begin{figure}[h!]
    \includegraphics[width = \linewidth]
    {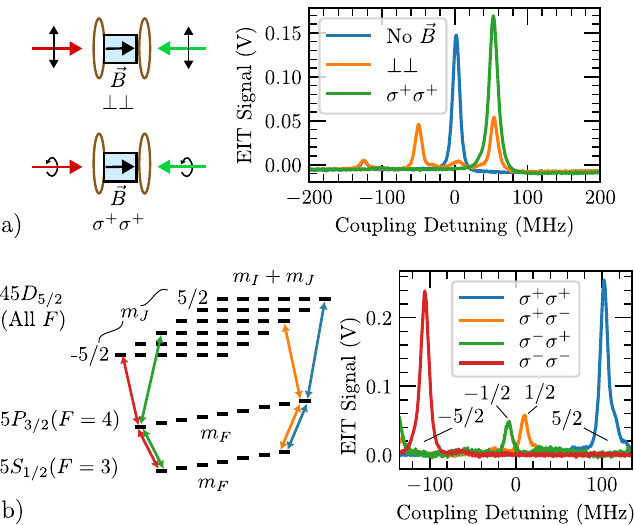}\\
    \caption{
    EIT in the presence of a large magnetic field using circularly polarized light. 
    a): Recovery of the measured EIT signal strength on $^{85}$Rb in a magnetic field of 1.39(1)~mT by using circular polarized light to populate $m_J = 5/2$ instead of linearly polarized light. The peak height exceeds that of the no-magnetic-field case. 
    b):
    Measured EIT signals on $^{85}$Rb in a magnetic field of 2.78(1)~mT under the four possible combinations of circularly polarized probe and coupling light with respect to the magnetic field axis. The polarizations are labeled as the probe polarization then the coupling polarization. Left: the predominant transition pathway for each polarization combination. Right: The resulting EIT peaks, with each peak labeled by its Rydberg $m_J$.
    }
    \label{fig:gainrecovery}
\end{figure}

The signal strength can be recovered by rotating the magnetic field to be aligned with the propagation axis of the lasers and using circularly polarized light. In this way, the entire population can be pumped to the $m_J = +5/2$ level, resulting in a slight increase in the resulting signal over the standard technique due to the higher transition dipole moments of the stretch transitions from the lower states to the Rydberg state compared to $\pi$ transitions. This is demonstrated in the ${5S_{1/2}(F=3)}\rightarrow{5P_{3/2}(F=4)}\rightarrow{45D_{5/2}}$ transition in $^{85}$Rb in Figure~\ref{fig:gainrecovery}~a).

Special care was taken to avoid polarization dependent effects in the measurement, such as using pickoff mirrors to overlap the probe and coupling beams at a slight angle rather than a dichroic, which would have polarization-dependent transmittance.
In addition, all four possible combinations of polarized light can populate various isolated $m_J$ levels, as depicted in Figure~\ref{fig:gainrecovery}~b).

\section{Conclusion}
With the recent introduction \cite{shi2023electric} of Zeeman shifts as a way to tune Autler-Townes electrometry,  we have made several key observations about this technique. First, we made sense of the shape of the EIT signals theoretically and demonstrated that the magnetic field defines the quantization axis. Next, we have demonstrated that isolating $m_J$ levels via Zeeman splitting leads to Autler-Townes splittings with a single transition and therefore a single dipole moment, thus making RF power calibration simpler. Finally, we have demonstrated in Figure~\ref{fig:gainrecovery}~a) that we can perform this technique with the same or a greater signal-to-noise ratio as without a magnetic field using circularly polarized light.

\section{Methods}
The magnetic field is applied with a proprietary set of Helmholtz coils with a coil diameter of 195~mm and of order 200~turns, with a nominal internal magnetic field of 1~mT/A but measured to be 0.926(1)~mT/A, varying by less than 10\% within the active region of the coil pair. The magnetic field values are reported as the current applied multiplied by this ratio. Background magnetic fields ($<0.1$~mT) are small compared to applied fields, so their effects are not included, but a 0.1~mT uncertainty is included in the error analysis. The Cs atoms are in a 25~mm diameter by 25~mm length cell, and the Rb atoms are in a 25~mm diameter by 75~mm length cell. The RF is applied with a broadband horn placed $\sim$54~cm from the cell.

\newpage
\begin{acknowledgments}
\vspace{-5mm}
\noindent This work was partially funded by the NIST-on-a-Chip (NOAC) Program.

\vspace{-5mm}
\section*{Conflict of Interest}
\vspace{-5mm}\noindent The authors have no conflicts of interests to disclose.

\vspace{-5mm}
\section*{Data Availability Statement}
\vspace{-5mm}\noindent The data relevant to the findings of this research project are available at \href{https://datapub.nist.gov/od/id/mds2-3102}{doi:10.18434/mds2-3102}.

\end{acknowledgments}

\bibliography{paper}

\begin{thebibliography}{35}%
\makeatletter
\providecommand \@ifxundefined [1]{%
 \@ifx{#1\undefined}
}%
\providecommand \@ifnum [1]{%
 \ifnum #1\expandafter \@firstoftwo
 \else \expandafter \@secondoftwo
 \fi
}%
\providecommand \@ifx [1]{%
 \ifx #1\expandafter \@firstoftwo
 \else \expandafter \@secondoftwo
 \fi
}%
\providecommand \natexlab [1]{#1}%
\providecommand \enquote  [1]{``#1''}%
\providecommand \bibnamefont  [1]{#1}%
\providecommand \bibfnamefont [1]{#1}%
\providecommand \citenamefont [1]{#1}%
\providecommand \href@noop [0]{\@secondoftwo}%
\providecommand \href [0]{\begingroup \@sanitize@url \@href}%
\providecommand \@href[1]{\@@startlink{#1}\@@href}%
\providecommand \@@href[1]{\endgroup#1\@@endlink}%
\providecommand \@sanitize@url [0]{\catcode `\\12\catcode `\$12\catcode `\&12\catcode `\#12\catcode `\^12\catcode `\_12\catcode `\%12\relax}%
\providecommand \@@startlink[1]{}%
\providecommand \@@endlink[0]{}%
\providecommand \url  [0]{\begingroup\@sanitize@url \@url }%
\providecommand \@url [1]{\endgroup\@href {#1}{\urlprefix }}%
\providecommand \urlprefix  [0]{URL }%
\providecommand \Eprint [0]{\href }%
\providecommand \doibase [0]{http://dx.doi.org/}%
\providecommand \selectlanguage [0]{\@gobble}%
\providecommand \bibinfo  [0]{\@secondoftwo}%
\providecommand \bibfield  [0]{\@secondoftwo}%
\providecommand \translation [1]{[#1]}%
\providecommand \BibitemOpen [0]{}%
\providecommand \bibitemStop [0]{}%
\providecommand \bibitemNoStop [0]{.\EOS\space}%
\providecommand \EOS [0]{\spacefactor3000\relax}%
\providecommand \BibitemShut  [1]{\csname bibitem#1\endcsname}%
\let\auto@bib@innerbib\@empty
\bibitem [{\citenamefont {Sedlacek}\ \emph {et~al.}(2012)\citenamefont {Sedlacek}, \citenamefont {Schwettmann}, \citenamefont {K{\"u}bler}, \citenamefont {L{\"o}w}, \citenamefont {Pfau},\ and\ \citenamefont {Shaffer}}]{sedlacek2012microwave}%
  \BibitemOpen
  \bibfield  {author} {\bibinfo {author} {\bibfnamefont {J.~A.}\ \bibnamefont {Sedlacek}}, \bibinfo {author} {\bibfnamefont {A.}~\bibnamefont {Schwettmann}}, \bibinfo {author} {\bibfnamefont {H.}~\bibnamefont {K{\"u}bler}}, \bibinfo {author} {\bibfnamefont {R.}~\bibnamefont {L{\"o}w}}, \bibinfo {author} {\bibfnamefont {T.}~\bibnamefont {Pfau}}, \ and\ \bibinfo {author} {\bibfnamefont {J.~P.}\ \bibnamefont {Shaffer}},\ }\href {\doibase 10.1038/nphys2423} {\bibfield  {journal} {\bibinfo  {journal} {Nature physics}\ }\textbf {\bibinfo {volume} {8}},\ \bibinfo {pages} {819} (\bibinfo {year} {2012})}\BibitemShut {NoStop}%
\bibitem [{\citenamefont {Holloway}\ \emph {et~al.}(2014)\citenamefont {Holloway}, \citenamefont {Gordon}, \citenamefont {Schwarzkopf}, \citenamefont {Anderson}, \citenamefont {Miller}, \citenamefont {Thaicharoen},\ and\ \citenamefont {Raithel}}]{holloway2014aps}%
  \BibitemOpen
  \bibfield  {author} {\bibinfo {author} {\bibfnamefont {C.~L.}\ \bibnamefont {Holloway}}, \bibinfo {author} {\bibfnamefont {J.}~\bibnamefont {Gordon}}, \bibinfo {author} {\bibfnamefont {A.}~\bibnamefont {Schwarzkopf}}, \bibinfo {author} {\bibfnamefont {D.}~\bibnamefont {Anderson}}, \bibinfo {author} {\bibfnamefont {S.}~\bibnamefont {Miller}}, \bibinfo {author} {\bibfnamefont {N.}~\bibnamefont {Thaicharoen}}, \ and\ \bibinfo {author} {\bibfnamefont {G.}~\bibnamefont {Raithel}},\ }\href {\doibase 10.1109/TAP.2014.2360208} {\bibfield  {journal} {\bibinfo  {journal} {IEEE Trans. on Antenna and Propag.}\ }\textbf {\bibinfo {volume} {62}},\ \bibinfo {pages} {6169} (\bibinfo {year} {2014})}\BibitemShut {NoStop}%
\bibitem [{\citenamefont {Artusio-Glimpse}\ \emph {et~al.}(2022)\citenamefont {Artusio-Glimpse}, \citenamefont {Simons}, \citenamefont {Prajapati},\ and\ \citenamefont {Holloway}}]{9748947}%
  \BibitemOpen
  \bibfield  {author} {\bibinfo {author} {\bibfnamefont {A.}~\bibnamefont {Artusio-Glimpse}}, \bibinfo {author} {\bibfnamefont {M.~T.}\ \bibnamefont {Simons}}, \bibinfo {author} {\bibfnamefont {N.}~\bibnamefont {Prajapati}}, \ and\ \bibinfo {author} {\bibfnamefont {C.~L.}\ \bibnamefont {Holloway}},\ }\href {\doibase 10.1109/MMM.2022.3148705} {\bibfield  {journal} {\bibinfo  {journal} {IEEE Microwave Magazine}\ }\textbf {\bibinfo {volume} {23}},\ \bibinfo {pages} {44} (\bibinfo {year} {2022})}\BibitemShut {NoStop}%
\bibitem [{\citenamefont {Holloway}\ \emph {et~al.}(2017)\citenamefont {Holloway}, \citenamefont {Simons}, \citenamefont {Gordon}, \citenamefont {Dienstfrey}, \citenamefont {Anderson},\ and\ \citenamefont {Raithel}}]{holloway2017electric}%
  \BibitemOpen
  \bibfield  {author} {\bibinfo {author} {\bibfnamefont {C.~L.}\ \bibnamefont {Holloway}}, \bibinfo {author} {\bibfnamefont {M.~T.}\ \bibnamefont {Simons}}, \bibinfo {author} {\bibfnamefont {J.~A.}\ \bibnamefont {Gordon}}, \bibinfo {author} {\bibfnamefont {A.}~\bibnamefont {Dienstfrey}}, \bibinfo {author} {\bibfnamefont {D.~A.}\ \bibnamefont {Anderson}}, \ and\ \bibinfo {author} {\bibfnamefont {G.}~\bibnamefont {Raithel}},\ }\href {\doibase 10.1063/1.4984201} {\bibfield  {journal} {\bibinfo  {journal} {Journal of Applied Physics}\ }\textbf {\bibinfo {volume} {121}} (\bibinfo {year} {2017}),\ 10.1063/1.4984201}\BibitemShut {NoStop}%
\bibitem [{\citenamefont {Meyer}\ \emph {et~al.}(2020)\citenamefont {Meyer}, \citenamefont {Castillo}, \citenamefont {Cox},\ and\ \citenamefont {Kunz}}]{meyer2020assessment}%
  \BibitemOpen
  \bibfield  {author} {\bibinfo {author} {\bibfnamefont {D.~H.}\ \bibnamefont {Meyer}}, \bibinfo {author} {\bibfnamefont {Z.~A.}\ \bibnamefont {Castillo}}, \bibinfo {author} {\bibfnamefont {K.~C.}\ \bibnamefont {Cox}}, \ and\ \bibinfo {author} {\bibfnamefont {P.~D.}\ \bibnamefont {Kunz}},\ }\href {\doibase 10.1088/1361-6455/ab6051} {\bibfield  {journal} {\bibinfo  {journal} {Journal of Physics B: Atomic, Molecular and Optical Physics}\ }\textbf {\bibinfo {volume} {53}},\ \bibinfo {pages} {034001} (\bibinfo {year} {2020})}\BibitemShut {NoStop}%
\bibitem [{\citenamefont {Fancher}\ \emph {et~al.}(2021)\citenamefont {Fancher}, \citenamefont {Scherer}, \citenamefont {John},\ and\ \citenamefont {Marlow}}]{fancher2021rydberg}%
  \BibitemOpen
  \bibfield  {author} {\bibinfo {author} {\bibfnamefont {C.~T.}\ \bibnamefont {Fancher}}, \bibinfo {author} {\bibfnamefont {D.~R.}\ \bibnamefont {Scherer}}, \bibinfo {author} {\bibfnamefont {M.~C.~S.}\ \bibnamefont {John}}, \ and\ \bibinfo {author} {\bibfnamefont {B.~L.~S.}\ \bibnamefont {Marlow}},\ }\href {\doibase 10.1109/TQE.2021.3065227} {\bibfield  {journal} {\bibinfo  {journal} {IEEE Transactions on Quantum Engineering}\ }\textbf {\bibinfo {volume} {2}},\ \bibinfo {pages} {1} (\bibinfo {year} {2021})}\BibitemShut {NoStop}%
\bibitem [{\citenamefont {Yuan}\ \emph {et~al.}(2023)\citenamefont {Yuan}, \citenamefont {Yang}, \citenamefont {Jing}, \citenamefont {Zhang}, \citenamefont {Jiao}, \citenamefont {Li}, \citenamefont {Zhang}, \citenamefont {Xiao},\ and\ \citenamefont {Jia}}]{yuan2023quantum}%
  \BibitemOpen
  \bibfield  {author} {\bibinfo {author} {\bibfnamefont {J.}~\bibnamefont {Yuan}}, \bibinfo {author} {\bibfnamefont {W.}~\bibnamefont {Yang}}, \bibinfo {author} {\bibfnamefont {M.}~\bibnamefont {Jing}}, \bibinfo {author} {\bibfnamefont {H.}~\bibnamefont {Zhang}}, \bibinfo {author} {\bibfnamefont {Y.}~\bibnamefont {Jiao}}, \bibinfo {author} {\bibfnamefont {W.}~\bibnamefont {Li}}, \bibinfo {author} {\bibfnamefont {L.}~\bibnamefont {Zhang}}, \bibinfo {author} {\bibfnamefont {L.}~\bibnamefont {Xiao}}, \ and\ \bibinfo {author} {\bibfnamefont {S.}~\bibnamefont {Jia}},\ }\href {\doibase 10.1088/1361-6633/acf22f} {\bibfield  {journal} {\bibinfo  {journal} {Reports on Progress in Physics}\ } (\bibinfo {year} {2023}),\ 10.1088/1361-6633/acf22f}\BibitemShut {NoStop}%
\bibitem [{\citenamefont {Liu}\ \emph {et~al.}(2023)\citenamefont {Liu}, \citenamefont {Zhang}, \citenamefont {Liu}, \citenamefont {Deng}, \citenamefont {Ding}, \citenamefont {Shi},\ and\ \citenamefont {Guo}}]{liu2023electric}%
  \BibitemOpen
  \bibfield  {author} {\bibinfo {author} {\bibfnamefont {B.}~\bibnamefont {Liu}}, \bibinfo {author} {\bibfnamefont {L.}~\bibnamefont {Zhang}}, \bibinfo {author} {\bibfnamefont {Z.}~\bibnamefont {Liu}}, \bibinfo {author} {\bibfnamefont {Z.}~\bibnamefont {Deng}}, \bibinfo {author} {\bibfnamefont {D.}~\bibnamefont {Ding}}, \bibinfo {author} {\bibfnamefont {B.}~\bibnamefont {Shi}}, \ and\ \bibinfo {author} {\bibfnamefont {G.}~\bibnamefont {Guo}},\ }\href {\doibase 10.23919/emsci.2022.0015} {\bibfield  {journal} {\bibinfo  {journal} {Electromagnetic Science}\ }\textbf {\bibinfo {volume} {1}},\ \bibinfo {pages} {1} (\bibinfo {year} {2023})}\BibitemShut {NoStop}%
\bibitem [{\citenamefont {{\v{S}}ibali{\'c}}\ \emph {et~al.}(2017)\citenamefont {{\v{S}}ibali{\'c}}, \citenamefont {Pritchard}, \citenamefont {Adams},\ and\ \citenamefont {Weatherill}}]{vsibalic2017arc}%
  \BibitemOpen
  \bibfield  {author} {\bibinfo {author} {\bibfnamefont {N.}~\bibnamefont {{\v{S}}ibali{\'c}}}, \bibinfo {author} {\bibfnamefont {J.~D.}\ \bibnamefont {Pritchard}}, \bibinfo {author} {\bibfnamefont {C.~S.}\ \bibnamefont {Adams}}, \ and\ \bibinfo {author} {\bibfnamefont {K.~J.}\ \bibnamefont {Weatherill}},\ }\href {\doibase 10.1016/j.cpc.2017.06.015} {\bibfield  {journal} {\bibinfo  {journal} {Computer Physics Communications}\ }\textbf {\bibinfo {volume} {220}},\ \bibinfo {pages} {319} (\bibinfo {year} {2017})}\BibitemShut {NoStop}%
\bibitem [{\citenamefont {Mohapatra}\ \emph {et~al.}(2007)\citenamefont {Mohapatra}, \citenamefont {Jackson},\ and\ \citenamefont {Adams}}]{mohapatra2007coherent}%
  \BibitemOpen
  \bibfield  {author} {\bibinfo {author} {\bibfnamefont {A.~K.}\ \bibnamefont {Mohapatra}}, \bibinfo {author} {\bibfnamefont {T.~R.}\ \bibnamefont {Jackson}}, \ and\ \bibinfo {author} {\bibfnamefont {C.~S.}\ \bibnamefont {Adams}},\ }\href {\doibase 10.1103/PhysRevLett.98.113003} {\bibfield  {journal} {\bibinfo  {journal} {Phys. Rev. Lett.}\ }\textbf {\bibinfo {volume} {98}},\ \bibinfo {pages} {113003} (\bibinfo {year} {2007})}\BibitemShut {NoStop}%
\bibitem [{\citenamefont {Simons}\ \emph {et~al.}(2016)\citenamefont {Simons}, \citenamefont {Gordon}, \citenamefont {Holloway}, \citenamefont {Anderson}, \citenamefont {Miller},\ and\ \citenamefont {Raithel}}]{simons2016using}%
  \BibitemOpen
  \bibfield  {author} {\bibinfo {author} {\bibfnamefont {M.~T.}\ \bibnamefont {Simons}}, \bibinfo {author} {\bibfnamefont {J.~A.}\ \bibnamefont {Gordon}}, \bibinfo {author} {\bibfnamefont {C.~L.}\ \bibnamefont {Holloway}}, \bibinfo {author} {\bibfnamefont {D.~A.}\ \bibnamefont {Anderson}}, \bibinfo {author} {\bibfnamefont {S.~A.}\ \bibnamefont {Miller}}, \ and\ \bibinfo {author} {\bibfnamefont {G.}~\bibnamefont {Raithel}},\ }\href {\doibase 10.1063/1.4947231} {\bibfield  {journal} {\bibinfo  {journal} {Applied Physics Letters}\ }\textbf {\bibinfo {volume} {108}} (\bibinfo {year} {2016}),\ 10.1063/1.4947231}\BibitemShut {NoStop}%
\bibitem [{\citenamefont {Simons}\ \emph {et~al.}(2021)\citenamefont {Simons}, \citenamefont {Artusio-Glimpse}, \citenamefont {Holloway}, \citenamefont {Imhof}, \citenamefont {Jefferts}, \citenamefont {Wyllie}, \citenamefont {Sawyer},\ and\ \citenamefont {Walker}}]{PhysRevA.104.032824}%
  \BibitemOpen
  \bibfield  {author} {\bibinfo {author} {\bibfnamefont {M.~T.}\ \bibnamefont {Simons}}, \bibinfo {author} {\bibfnamefont {A.~B.}\ \bibnamefont {Artusio-Glimpse}}, \bibinfo {author} {\bibfnamefont {C.~L.}\ \bibnamefont {Holloway}}, \bibinfo {author} {\bibfnamefont {E.}~\bibnamefont {Imhof}}, \bibinfo {author} {\bibfnamefont {S.~R.}\ \bibnamefont {Jefferts}}, \bibinfo {author} {\bibfnamefont {R.}~\bibnamefont {Wyllie}}, \bibinfo {author} {\bibfnamefont {B.~C.}\ \bibnamefont {Sawyer}}, \ and\ \bibinfo {author} {\bibfnamefont {T.~G.}\ \bibnamefont {Walker}},\ }\href {\doibase 10.1103/PhysRevA.104.032824} {\bibfield  {journal} {\bibinfo  {journal} {Phys. Rev. A}\ }\textbf {\bibinfo {volume} {104}},\ \bibinfo {pages} {032824} (\bibinfo {year} {2021})}\BibitemShut {NoStop}%
\bibitem [{\citenamefont {Anderson}\ and\ \citenamefont {Raithel}(2017)}]{10.1063/1.4996234}%
  \BibitemOpen
  \bibfield  {author} {\bibinfo {author} {\bibfnamefont {D.~A.}\ \bibnamefont {Anderson}}\ and\ \bibinfo {author} {\bibfnamefont {G.}~\bibnamefont {Raithel}},\ }\href {\doibase 10.1063/1.4996234} {\bibfield  {journal} {\bibinfo  {journal} {Applied Physics Letters}\ }\textbf {\bibinfo {volume} {111}},\ \bibinfo {pages} {053504} (\bibinfo {year} {2017})}\BibitemShut {NoStop}%
\bibitem [{\citenamefont {Hu}\ \emph {et~al.}(2022)\citenamefont {Hu}, \citenamefont {Li}, \citenamefont {Song}, \citenamefont {Bai}, \citenamefont {Jiao}, \citenamefont {Zhao},\ and\ \citenamefont {Jia}}]{hu2022continuously}%
  \BibitemOpen
  \bibfield  {author} {\bibinfo {author} {\bibfnamefont {J.}~\bibnamefont {Hu}}, \bibinfo {author} {\bibfnamefont {H.}~\bibnamefont {Li}}, \bibinfo {author} {\bibfnamefont {R.}~\bibnamefont {Song}}, \bibinfo {author} {\bibfnamefont {J.}~\bibnamefont {Bai}}, \bibinfo {author} {\bibfnamefont {Y.}~\bibnamefont {Jiao}}, \bibinfo {author} {\bibfnamefont {J.}~\bibnamefont {Zhao}}, \ and\ \bibinfo {author} {\bibfnamefont {S.}~\bibnamefont {Jia}},\ }\href {\doibase 10.1063/5.0086357} {\bibfield  {journal} {\bibinfo  {journal} {Applied Physics Letters}\ }\textbf {\bibinfo {volume} {121}} (\bibinfo {year} {2022}),\ 10.1063/5.0086357}\BibitemShut {NoStop}%
\bibitem [{\citenamefont {Meyer}\ \emph {et~al.}(2021)\citenamefont {Meyer}, \citenamefont {Kunz},\ and\ \citenamefont {Cox}}]{meyer2021waveguide}%
  \BibitemOpen
  \bibfield  {author} {\bibinfo {author} {\bibfnamefont {D.~H.}\ \bibnamefont {Meyer}}, \bibinfo {author} {\bibfnamefont {P.~D.}\ \bibnamefont {Kunz}}, \ and\ \bibinfo {author} {\bibfnamefont {K.~C.}\ \bibnamefont {Cox}},\ }\href {\doibase 10.1103/PhysRevApplied.15.014053} {\bibfield  {journal} {\bibinfo  {journal} {Physical review applied}\ }\textbf {\bibinfo {volume} {15}},\ \bibinfo {pages} {014053} (\bibinfo {year} {2021})}\BibitemShut {NoStop}%
\bibitem [{\citenamefont {Osterwalder}\ and\ \citenamefont {Merkt}(1999)}]{osterwalder1999using}%
  \BibitemOpen
  \bibfield  {author} {\bibinfo {author} {\bibfnamefont {A.}~\bibnamefont {Osterwalder}}\ and\ \bibinfo {author} {\bibfnamefont {F.}~\bibnamefont {Merkt}},\ }\href {\doibase 10.1103/PhysRevLett.82.1831} {\bibfield  {journal} {\bibinfo  {journal} {Physical review letters}\ }\textbf {\bibinfo {volume} {82}},\ \bibinfo {pages} {1831} (\bibinfo {year} {1999})}\BibitemShut {NoStop}%
\bibitem [{\citenamefont {Ma}\ \emph {et~al.}(2022)\citenamefont {Ma}, \citenamefont {Viray}, \citenamefont {Anderson},\ and\ \citenamefont {Raithel}}]{ma2022measurement}%
  \BibitemOpen
  \bibfield  {author} {\bibinfo {author} {\bibfnamefont {L.}~\bibnamefont {Ma}}, \bibinfo {author} {\bibfnamefont {M.~A.}\ \bibnamefont {Viray}}, \bibinfo {author} {\bibfnamefont {D.~A.}\ \bibnamefont {Anderson}}, \ and\ \bibinfo {author} {\bibfnamefont {G.}~\bibnamefont {Raithel}},\ }\href {\doibase 10.1103/PhysRevApplied.18.024001} {\bibfield  {journal} {\bibinfo  {journal} {Physical Review Applied}\ }\textbf {\bibinfo {volume} {18}},\ \bibinfo {pages} {024001} (\bibinfo {year} {2022})}\BibitemShut {NoStop}%
\bibitem [{\citenamefont {Holloway}\ \emph {et~al.}(2022)\citenamefont {Holloway}, \citenamefont {Prajapati}, \citenamefont {Sherman}, \citenamefont {R{\"u}fenacht}, \citenamefont {Artusio-Glimpse}, \citenamefont {Simons}, \citenamefont {Robinson}, \citenamefont {La~Mantia},\ and\ \citenamefont {Norrgard}}]{holloway2022electromagnetically}%
  \BibitemOpen
  \bibfield  {author} {\bibinfo {author} {\bibfnamefont {C.~L.}\ \bibnamefont {Holloway}}, \bibinfo {author} {\bibfnamefont {N.}~\bibnamefont {Prajapati}}, \bibinfo {author} {\bibfnamefont {J.~A.}\ \bibnamefont {Sherman}}, \bibinfo {author} {\bibfnamefont {A.}~\bibnamefont {R{\"u}fenacht}}, \bibinfo {author} {\bibfnamefont {A.~B.}\ \bibnamefont {Artusio-Glimpse}}, \bibinfo {author} {\bibfnamefont {M.~T.}\ \bibnamefont {Simons}}, \bibinfo {author} {\bibfnamefont {A.~K.}\ \bibnamefont {Robinson}}, \bibinfo {author} {\bibfnamefont {D.~S.}\ \bibnamefont {La~Mantia}}, \ and\ \bibinfo {author} {\bibfnamefont {E.~B.}\ \bibnamefont {Norrgard}},\ }\href {\doibase 10.1116/5.0097746} {\bibfield  {journal} {\bibinfo  {journal} {AVS Quantum Science}\ }\textbf {\bibinfo {volume} {4}} (\bibinfo {year} {2022}),\ 10.1116/5.0097746}\BibitemShut {NoStop}%
\bibitem [{\citenamefont {Duspayev}\ \emph {et~al.}(2023)\citenamefont {Duspayev}, \citenamefont {Cardman}, \citenamefont {Anderson},\ and\ \citenamefont {Raithel}}]{duspayev2023high}%
  \BibitemOpen
  \bibfield  {author} {\bibinfo {author} {\bibfnamefont {A.}~\bibnamefont {Duspayev}}, \bibinfo {author} {\bibfnamefont {R.}~\bibnamefont {Cardman}}, \bibinfo {author} {\bibfnamefont {D.~A.}\ \bibnamefont {Anderson}}, \ and\ \bibinfo {author} {\bibfnamefont {G.}~\bibnamefont {Raithel}},\ }\href {\doibase 10.48550/arXiv.2310.10542} {\bibfield  {journal} {\bibinfo  {journal} {arXiv preprint arXiv:2310.10542}\ } (\bibinfo {year} {2023}),\ 10.48550/arXiv.2310.10542}\BibitemShut {NoStop}%
\bibitem [{\citenamefont {Noel}\ \emph {et~al.}(1998)\citenamefont {Noel}, \citenamefont {Griffith},\ and\ \citenamefont {Gallagher}}]{noel1998frequency}%
  \BibitemOpen
  \bibfield  {author} {\bibinfo {author} {\bibfnamefont {M.~W.}\ \bibnamefont {Noel}}, \bibinfo {author} {\bibfnamefont {W.~M.}\ \bibnamefont {Griffith}}, \ and\ \bibinfo {author} {\bibfnamefont {T.~F.}\ \bibnamefont {Gallagher}},\ }\href {\doibase 10.1103/PhysRevA.58.2265} {\bibfield  {journal} {\bibinfo  {journal} {Physical Review A}\ }\textbf {\bibinfo {volume} {58}},\ \bibinfo {pages} {2265} (\bibinfo {year} {1998})}\BibitemShut {NoStop}%
\bibitem [{\citenamefont {Bohlouli-Zanjani}\ \emph {et~al.}(2007)\citenamefont {Bohlouli-Zanjani}, \citenamefont {Petrus},\ and\ \citenamefont {Martin}}]{bohlouli2007enhancement}%
  \BibitemOpen
  \bibfield  {author} {\bibinfo {author} {\bibfnamefont {P.}~\bibnamefont {Bohlouli-Zanjani}}, \bibinfo {author} {\bibfnamefont {J.~A.}\ \bibnamefont {Petrus}}, \ and\ \bibinfo {author} {\bibfnamefont {J.~D.~D.}\ \bibnamefont {Martin}},\ }\href {\doibase 10.1103/PhysRevLett.98.203005} {\bibfield  {journal} {\bibinfo  {journal} {Physical review letters}\ }\textbf {\bibinfo {volume} {98}},\ \bibinfo {pages} {203005} (\bibinfo {year} {2007})}\BibitemShut {NoStop}%
\bibitem [{\citenamefont {Berweger}\ \emph {et~al.}(2023)\citenamefont {Berweger}, \citenamefont {Prajapati}, \citenamefont {Artusio-Glimpse}, \citenamefont {Rotunno}, \citenamefont {Brown}, \citenamefont {Holloway}, \citenamefont {Simons}, \citenamefont {Imhof}, \citenamefont {Jefferts}, \citenamefont {Kayim} \emph {et~al.}}]{berweger2023rydberg}%
  \BibitemOpen
  \bibfield  {author} {\bibinfo {author} {\bibfnamefont {S.}~\bibnamefont {Berweger}}, \bibinfo {author} {\bibfnamefont {N.}~\bibnamefont {Prajapati}}, \bibinfo {author} {\bibfnamefont {A.~B.}\ \bibnamefont {Artusio-Glimpse}}, \bibinfo {author} {\bibfnamefont {A.~P.}\ \bibnamefont {Rotunno}}, \bibinfo {author} {\bibfnamefont {R.}~\bibnamefont {Brown}}, \bibinfo {author} {\bibfnamefont {C.~L.}\ \bibnamefont {Holloway}}, \bibinfo {author} {\bibfnamefont {M.~T.}\ \bibnamefont {Simons}}, \bibinfo {author} {\bibfnamefont {E.}~\bibnamefont {Imhof}}, \bibinfo {author} {\bibfnamefont {S.~R.}\ \bibnamefont {Jefferts}}, \bibinfo {author} {\bibfnamefont {B.~N.}\ \bibnamefont {Kayim}},  \emph {et~al.},\ }\href {\doibase 10.1103/PhysRevApplied.19.044049} {\bibfield  {journal} {\bibinfo  {journal} {Physical Review Applied}\ }\textbf {\bibinfo {volume} {19}},\ \bibinfo {pages} {044049} (\bibinfo {year} {2023})}\BibitemShut {NoStop}%
\bibitem [{\citenamefont {Rotunno}\ \emph {et~al.}(2023{\natexlab{a}})\citenamefont {Rotunno}, \citenamefont {Berweger}, \citenamefont {Prajapati}, \citenamefont {Simons}, \citenamefont {Artusio-Glimpse}, \citenamefont {Holloway}, \citenamefont {Jayaseelan}, \citenamefont {Potvliege},\ and\ \citenamefont {Adams}}]{rotunno2023detection}%
  \BibitemOpen
  \bibfield  {author} {\bibinfo {author} {\bibfnamefont {A.~P.}\ \bibnamefont {Rotunno}}, \bibinfo {author} {\bibfnamefont {S.}~\bibnamefont {Berweger}}, \bibinfo {author} {\bibfnamefont {N.}~\bibnamefont {Prajapati}}, \bibinfo {author} {\bibfnamefont {M.~T.}\ \bibnamefont {Simons}}, \bibinfo {author} {\bibfnamefont {A.~B.}\ \bibnamefont {Artusio-Glimpse}}, \bibinfo {author} {\bibfnamefont {C.~L.}\ \bibnamefont {Holloway}}, \bibinfo {author} {\bibfnamefont {M.}~\bibnamefont {Jayaseelan}}, \bibinfo {author} {\bibfnamefont {R.}~\bibnamefont {Potvliege}}, \ and\ \bibinfo {author} {\bibfnamefont {C.}~\bibnamefont {Adams}},\ }\href {\doibase 10.1063/5.0162101} {\bibfield  {journal} {\bibinfo  {journal} {Journal of Applied Physics}\ }\textbf {\bibinfo {volume} {134}} (\bibinfo {year} {2023}{\natexlab{a}}),\ 10.1063/5.0162101}\BibitemShut {NoStop}%
\bibitem [{\citenamefont {Bao}\ \emph {et~al.}(2016)\citenamefont {Bao}, \citenamefont {Zhang}, \citenamefont {Zhou}, \citenamefont {Zhang}, \citenamefont {Zhao}, \citenamefont {Xiao},\ and\ \citenamefont {Jia}}]{bao2016polarization}%
  \BibitemOpen
  \bibfield  {author} {\bibinfo {author} {\bibfnamefont {S.}~\bibnamefont {Bao}}, \bibinfo {author} {\bibfnamefont {H.}~\bibnamefont {Zhang}}, \bibinfo {author} {\bibfnamefont {J.}~\bibnamefont {Zhou}}, \bibinfo {author} {\bibfnamefont {L.}~\bibnamefont {Zhang}}, \bibinfo {author} {\bibfnamefont {J.}~\bibnamefont {Zhao}}, \bibinfo {author} {\bibfnamefont {L.}~\bibnamefont {Xiao}}, \ and\ \bibinfo {author} {\bibfnamefont {S.}~\bibnamefont {Jia}},\ }\href {\doibase 10.1103/PhysRevA.94.043822} {\bibfield  {journal} {\bibinfo  {journal} {Physical Review A}\ }\textbf {\bibinfo {volume} {94}},\ \bibinfo {pages} {043822} (\bibinfo {year} {2016})}\BibitemShut {NoStop}%
\bibitem [{\citenamefont {Naber}\ \emph {et~al.}(2017)\citenamefont {Naber}, \citenamefont {Tauschinsky}, \citenamefont {van Linden van~den Heuvell},\ and\ \citenamefont {Spreeuw}}]{naber2017electromagnetically}%
  \BibitemOpen
  \bibfield  {author} {\bibinfo {author} {\bibfnamefont {J.}~\bibnamefont {Naber}}, \bibinfo {author} {\bibfnamefont {A.}~\bibnamefont {Tauschinsky}}, \bibinfo {author} {\bibfnamefont {B.}~\bibnamefont {van Linden van~den Heuvell}}, \ and\ \bibinfo {author} {\bibfnamefont {R.}~\bibnamefont {Spreeuw}},\ }\href {\doibase 10.21468/SciPostPhys.2.2.015} {\bibfield  {journal} {\bibinfo  {journal} {SciPost Physics}\ }\textbf {\bibinfo {volume} {2}},\ \bibinfo {pages} {015} (\bibinfo {year} {2017})}\BibitemShut {NoStop}%
\bibitem [{\citenamefont {Shi}\ \emph {et~al.}(2023)\citenamefont {Shi}, \citenamefont {Li}, \citenamefont {Ouyang}, \citenamefont {Ren}, \citenamefont {Li}, \citenamefont {Cao}, \citenamefont {Xue},\ and\ \citenamefont {Shi}}]{shi2023electric}%
  \BibitemOpen
  \bibfield  {author} {\bibinfo {author} {\bibfnamefont {Y.}~\bibnamefont {Shi}}, \bibinfo {author} {\bibfnamefont {C.}~\bibnamefont {Li}}, \bibinfo {author} {\bibfnamefont {K.}~\bibnamefont {Ouyang}}, \bibinfo {author} {\bibfnamefont {W.}~\bibnamefont {Ren}}, \bibinfo {author} {\bibfnamefont {W.}~\bibnamefont {Li}}, \bibinfo {author} {\bibfnamefont {M.}~\bibnamefont {Cao}}, \bibinfo {author} {\bibfnamefont {Z.}~\bibnamefont {Xue}}, \ and\ \bibinfo {author} {\bibfnamefont {M.}~\bibnamefont {Shi}},\ }\href {\doibase 10.1364/OE.501647} {\bibfield  {journal} {\bibinfo  {journal} {Opt. Express}\ }\textbf {\bibinfo {volume} {31}},\ \bibinfo {pages} {36255} (\bibinfo {year} {2023})}\BibitemShut {NoStop}%
\bibitem [{\citenamefont {Li}\ \emph {et~al.}(2023)\citenamefont {Li}, \citenamefont {Cui}, \citenamefont {Hao}, \citenamefont {Zhou}, \citenamefont {Wang}, \citenamefont {Jia}, \citenamefont {Zhang}, \citenamefont {Xie},\ and\ \citenamefont {Zhong}}]{li2023magnetic}%
  \BibitemOpen
  \bibfield  {author} {\bibinfo {author} {\bibfnamefont {X.}~\bibnamefont {Li}}, \bibinfo {author} {\bibfnamefont {Y.}~\bibnamefont {Cui}}, \bibinfo {author} {\bibfnamefont {J.}~\bibnamefont {Hao}}, \bibinfo {author} {\bibfnamefont {F.}~\bibnamefont {Zhou}}, \bibinfo {author} {\bibfnamefont {Y.}~\bibnamefont {Wang}}, \bibinfo {author} {\bibfnamefont {F.}~\bibnamefont {Jia}}, \bibinfo {author} {\bibfnamefont {J.}~\bibnamefont {Zhang}}, \bibinfo {author} {\bibfnamefont {F.}~\bibnamefont {Xie}}, \ and\ \bibinfo {author} {\bibfnamefont {Z.}~\bibnamefont {Zhong}},\ }\href {\doibase 10.1364/OE.505488} {\bibfield  {journal} {\bibinfo  {journal} {Opt. Express}\ }\textbf {\bibinfo {volume} {31}},\ \bibinfo {pages} {38165} (\bibinfo {year} {2023})}\BibitemShut {NoStop}%
\bibitem [{\citenamefont {Sedlacek}\ \emph {et~al.}(2013)\citenamefont {Sedlacek}, \citenamefont {Schwettmann}, \citenamefont {K{\"u}bler},\ and\ \citenamefont {Shaffer}}]{sedlacek2013atom}%
  \BibitemOpen
  \bibfield  {author} {\bibinfo {author} {\bibfnamefont {J.~A.}\ \bibnamefont {Sedlacek}}, \bibinfo {author} {\bibfnamefont {A.}~\bibnamefont {Schwettmann}}, \bibinfo {author} {\bibfnamefont {H.}~\bibnamefont {K{\"u}bler}}, \ and\ \bibinfo {author} {\bibfnamefont {J.~P.}\ \bibnamefont {Shaffer}},\ }\href {\doibase 10.1103/PhysRevLett.111.063001} {\bibfield  {journal} {\bibinfo  {journal} {Physical review letters}\ }\textbf {\bibinfo {volume} {111}},\ \bibinfo {pages} {063001} (\bibinfo {year} {2013})}\BibitemShut {NoStop}%
\bibitem [{\citenamefont {Wang}\ \emph {et~al.}(2023)\citenamefont {Wang}, \citenamefont {Jia}, \citenamefont {Hao}, \citenamefont {Cui}, \citenamefont {Zhou}, \citenamefont {Liu}, \citenamefont {Mei}, \citenamefont {Yu}, \citenamefont {Liu}, \citenamefont {Zhang} \emph {et~al.}}]{wang2023precise}%
  \BibitemOpen
  \bibfield  {author} {\bibinfo {author} {\bibfnamefont {Y.}~\bibnamefont {Wang}}, \bibinfo {author} {\bibfnamefont {F.}~\bibnamefont {Jia}}, \bibinfo {author} {\bibfnamefont {J.}~\bibnamefont {Hao}}, \bibinfo {author} {\bibfnamefont {Y.}~\bibnamefont {Cui}}, \bibinfo {author} {\bibfnamefont {F.}~\bibnamefont {Zhou}}, \bibinfo {author} {\bibfnamefont {X.}~\bibnamefont {Liu}}, \bibinfo {author} {\bibfnamefont {J.}~\bibnamefont {Mei}}, \bibinfo {author} {\bibfnamefont {Y.}~\bibnamefont {Yu}}, \bibinfo {author} {\bibfnamefont {Y.}~\bibnamefont {Liu}}, \bibinfo {author} {\bibfnamefont {J.}~\bibnamefont {Zhang}},  \emph {et~al.},\ }\href {\doibase 10.1364/OE.485662} {\bibfield  {journal} {\bibinfo  {journal} {Optics Express}\ }\textbf {\bibinfo {volume} {31}},\ \bibinfo {pages} {10449} (\bibinfo {year} {2023})}\BibitemShut {NoStop}%
\bibitem [{\citenamefont {Anderson}\ \emph {et~al.}(2014)\citenamefont {Anderson}, \citenamefont {Schwarzkopf}, \citenamefont {Miller}, \citenamefont {Thaicharoen}, \citenamefont {Raithel}, \citenamefont {Gordon},\ and\ \citenamefont {Holloway}}]{r13}%
  \BibitemOpen
  \bibfield  {author} {\bibinfo {author} {\bibfnamefont {D.~A.}\ \bibnamefont {Anderson}}, \bibinfo {author} {\bibfnamefont {A.}~\bibnamefont {Schwarzkopf}}, \bibinfo {author} {\bibfnamefont {S.~A.}\ \bibnamefont {Miller}}, \bibinfo {author} {\bibfnamefont {N.}~\bibnamefont {Thaicharoen}}, \bibinfo {author} {\bibfnamefont {G.}~\bibnamefont {Raithel}}, \bibinfo {author} {\bibfnamefont {J.~A.}\ \bibnamefont {Gordon}}, \ and\ \bibinfo {author} {\bibfnamefont {C.~L.}\ \bibnamefont {Holloway}},\ }\href {\doibase 10.1103/PhysRevA.90.043419} {\bibfield  {journal} {\bibinfo  {journal} {Phys. Rev. A}\ }\textbf {\bibinfo {volume} {90}},\ \bibinfo {pages} {043419} (\bibinfo {year} {2014})}\BibitemShut {NoStop}%
\bibitem [{\citenamefont {Anderson}\ \emph {et~al.}(2016)\citenamefont {Anderson}, \citenamefont {Miller}, \citenamefont {Raithel}, \citenamefont {Gordon}, \citenamefont {Butler},\ and\ \citenamefont {Holloway}}]{r12}%
  \BibitemOpen
  \bibfield  {author} {\bibinfo {author} {\bibfnamefont {D.~A.}\ \bibnamefont {Anderson}}, \bibinfo {author} {\bibfnamefont {S.~A.}\ \bibnamefont {Miller}}, \bibinfo {author} {\bibfnamefont {G.}~\bibnamefont {Raithel}}, \bibinfo {author} {\bibfnamefont {J.~A.}\ \bibnamefont {Gordon}}, \bibinfo {author} {\bibfnamefont {M.~L.}\ \bibnamefont {Butler}}, \ and\ \bibinfo {author} {\bibfnamefont {C.~L.}\ \bibnamefont {Holloway}},\ }\href {\doibase 10.1103/PhysRevApplied.5.034003} {\bibfield  {journal} {\bibinfo  {journal} {Phys. Rev. Appl.}\ }\textbf {\bibinfo {volume} {5}},\ \bibinfo {pages} {034003} (\bibinfo {year} {2016})}\BibitemShut {NoStop}%
\bibitem [{\citenamefont {Rotunno}\ \emph {et~al.}(2023{\natexlab{b}})\citenamefont {Rotunno}, \citenamefont {Holloway}, \citenamefont {Prajapati}, \citenamefont {Berweger}, \citenamefont {Artusio-Glimpse}, \citenamefont {Brown}, \citenamefont {Simons}, \citenamefont {Robinson}, \citenamefont {Kayim}, \citenamefont {Viray} \emph {et~al.}}]{rotunno2023investigating}%
  \BibitemOpen
  \bibfield  {author} {\bibinfo {author} {\bibfnamefont {A.~P.}\ \bibnamefont {Rotunno}}, \bibinfo {author} {\bibfnamefont {C.~L.}\ \bibnamefont {Holloway}}, \bibinfo {author} {\bibfnamefont {N.}~\bibnamefont {Prajapati}}, \bibinfo {author} {\bibfnamefont {S.}~\bibnamefont {Berweger}}, \bibinfo {author} {\bibfnamefont {A.~B.}\ \bibnamefont {Artusio-Glimpse}}, \bibinfo {author} {\bibfnamefont {R.}~\bibnamefont {Brown}}, \bibinfo {author} {\bibfnamefont {M.}~\bibnamefont {Simons}}, \bibinfo {author} {\bibfnamefont {A.~K.}\ \bibnamefont {Robinson}}, \bibinfo {author} {\bibfnamefont {B.~N.}\ \bibnamefont {Kayim}}, \bibinfo {author} {\bibfnamefont {M.~A.}\ \bibnamefont {Viray}},  \emph {et~al.},\ }\href {\doibase 10.1063/5.0161213} {\bibfield  {journal} {\bibinfo  {journal} {Journal of Applied Physics}\ }\textbf {\bibinfo {volume} {134}} (\bibinfo {year} {2023}{\natexlab{b}}),\ 10.1063/5.0161213}\BibitemShut {NoStop}%
\bibitem [{\citenamefont {Jia}\ \emph {et~al.}(2021)\citenamefont {Jia}, \citenamefont {Zhang}, \citenamefont {Liu}, \citenamefont {Mei}, \citenamefont {Yu}, \citenamefont {Lin}, \citenamefont {Dong}, \citenamefont {Liu}, \citenamefont {Zhang}, \citenamefont {Xie},\ and\ \citenamefont {Zhong}}]{ZeemanDipoleTreatment}%
  \BibitemOpen
  \bibfield  {author} {\bibinfo {author} {\bibfnamefont {F.-D.}\ \bibnamefont {Jia}}, \bibinfo {author} {\bibfnamefont {H.-Y.}\ \bibnamefont {Zhang}}, \bibinfo {author} {\bibfnamefont {X.-B.}\ \bibnamefont {Liu}}, \bibinfo {author} {\bibfnamefont {J.}~\bibnamefont {Mei}}, \bibinfo {author} {\bibfnamefont {Y.-H.}\ \bibnamefont {Yu}}, \bibinfo {author} {\bibfnamefont {Z.-Q.}\ \bibnamefont {Lin}}, \bibinfo {author} {\bibfnamefont {H.-Y.}\ \bibnamefont {Dong}}, \bibinfo {author} {\bibfnamefont {Y.}~\bibnamefont {Liu}}, \bibinfo {author} {\bibfnamefont {J.}~\bibnamefont {Zhang}}, \bibinfo {author} {\bibfnamefont {F.}~\bibnamefont {Xie}}, \ and\ \bibinfo {author} {\bibfnamefont {Z.-P.}\ \bibnamefont {Zhong}},\ }\href {\doibase 10.1088/1361-6455/ac1b66} {\bibfield  {journal} {\bibinfo  {journal} {Journal of Physics B: Atomic, Molecular and Optical Physics}\ }\textbf {\bibinfo {volume} {54}},\ \bibinfo {pages} {165501} (\bibinfo {year} {2021})}\BibitemShut {NoStop}%
\bibitem [{\citenamefont {Steck}(2003)}]{steck2003cesium}%
  \BibitemOpen
  \bibfield  {author} {\bibinfo {author} {\bibfnamefont {D.~A.}\ \bibnamefont {Steck}},\ }\href {https://steck.us/alkalidata/cesiumnumbers.pdf} {\enquote {\bibinfo {title} {Cesium {D} line data},}\ } (\bibinfo {year} {2003})\BibitemShut {NoStop}%
\bibitem [{\citenamefont {Miller}\ \emph {et~al.}(2023)\citenamefont {Miller}, \citenamefont {Meyer}, \citenamefont {Virtanen}, \citenamefont {O'Brien},\ and\ \citenamefont {Cox}}]{miller2023rydiqule}%
  \BibitemOpen
  \bibfield  {author} {\bibinfo {author} {\bibfnamefont {B.~N.}\ \bibnamefont {Miller}}, \bibinfo {author} {\bibfnamefont {D.~H.}\ \bibnamefont {Meyer}}, \bibinfo {author} {\bibfnamefont {T.}~\bibnamefont {Virtanen}}, \bibinfo {author} {\bibfnamefont {C.~M.}\ \bibnamefont {O'Brien}}, \ and\ \bibinfo {author} {\bibfnamefont {K.~C.}\ \bibnamefont {Cox}},\ }\href {\doibase 10.1016/j.cpc.2023.108952} {\bibfield  {journal} {\bibinfo  {journal} {Computer Physics Communications}\ ,\ \bibinfo {pages} {108952}} (\bibinfo {year} {2023})}\BibitemShut {NoStop}%
\end{thebibliography}%

\end{document}